\documentclass[conference]{IEEEtran}
\IEEEoverridecommandlockouts
\usepackage{cite}
\usepackage{amsmath,amssymb,amsfonts}
\usepackage{graphicx}
\usepackage{subcaption}
\usepackage{caption}
\usepackage{textcomp}
\usepackage{xcolor}
\usepackage{bm}
\usepackage{tabularx}
\usepackage[square, comma, sort&compress, numbers]{natbib}
\usepackage{float}  % 提供更多的浮动选项
\usepackage{diagbox} % 导入 diagbox 包
\usepackage{hyperref}
\usepackage{algorithm}
\usepackage{algpseudocode}
\def\BibTeX{{\rm B\kern-.05em{\sc i\kern-.025em b}\kern-.08em
    T\kern-.1667em\lower.7ex\hbox{E}\kern-.125emX}}
\begin{document}

\title{Generative CKM Construction using Partially Observed Data with Diffusion Model\\
	% \thanks{This work was supported by the National Key R\&D Program of China with Grant number 2019YFB1803400.}
}

\author{\IEEEauthorblockN{Shen Fu$^{*}$, Zijian Wu$^{*}$, Di Wu$^{*\dag}$, and Yong Zeng$^{*\dag }$}
	\IEEEauthorblockA{$^{*}$National Mobile Communications Research Laboratory, Southeast University, Nanjing 210096, China}
	\IEEEauthorblockA{$^{\dag}$Purple Mountain Laboratories, Nanjing 211111, China}
	\IEEEauthorblockA{sfu@seu.edu.cn, wuzijian@seu.edu.cn, studywudi@seu.edu.cn, yong\_zeng@seu.edu.cn}
}

\maketitle

\begin{abstract}
	% The evolution towards 6G wireless communication systems introduces significant advancements, such as denser infrastructure, larger antenna arrays, wider bandwidths, and heightened intelligence. However, these enhancements also present some challenges, particularly in real-time channel state information (CSI) acquisition. Channel knowledge map (CKM) is a site-specific database, tagged with the locations of transmitters/receivers, which contains useful channel information, facilitating and even obviating the need for real-time CSI acquisition. Due to the high cost of channel data measurements, our access to channel information for each geographical region is extremely restricted, posing a critical challenge for CKM construction that we need to recover the complete channel data from the sparse data sets. In this paper, we consider the CKMs as pictures. By leveraging generative Artificial Intelligence (AI) techniques, specifically diffusion models, we propose an innovative method for reconstructing incomplete CKM images. It is illustrated in simulations that complete CKM images are generated, which supports the deployment of the communication infrastructure.
    Channel knowledge map (CKM) is a promising technique that enables environment-aware wireless networks by utilizing location-specific channel prior information to improve communication and sensing performance. A fundamental problem for CKM construction is how to utilize partially observed channel knowledge data to reconstruct a complete CKM for all possible locations of interest. This problem resembles the long-standing ill-posed inverse problem, which tries to infer from a set of limited observations the cause factors that produced them. By utilizing the recent advances of solving inverse problems with generative artificial intelligence (AI), in this paper, we propose generative CKM construction method using partially observed data by solving inverse problems with diffusion models. Simulation results show that the proposed method significantly improves the performance of CKM construction compared with benchmarking schemes.
\end{abstract}


\section{Introduction}
% The pursuit of improved efficiency and reliability in the rapidly evolving field of wireless communications has prompted researchers to explore innovative paradigms that go beyond conventional approaches.
The ever-increasing demand for efficient and reliable wireless communication drives researchers to explore innovative approaches, such as extremely large-scale multiple-input multiple-output (XL-MIMO) \cite{lhq} and millimeter-wave (mmWave) communication.
However, obtaining accurate and timely channel state information (CSI) in complex communication environment remains challenging \cite{csi1}.
% However, obtaining accurate and timely channel state information (CSI) remains a significant challenge \cite{csi1}.
% An important challenge is the acquisition of channel state information (CSI) \cite{csi2}.
To address this issue, \cite{ckm} proposed the concept of channel knowledge map (CKM), which is a site-specific database, tagged with the locations of the transmitters and receivers.
By enabling environment-aware communications and sensing, CKM facilitates real-time CSI acquisition, thereby promising better performance than conventional environment-unaware communications which heavily rely on real-time channel training.
The significance of CKM is especially appealing for scenarios where real-time channel training is either extremely costly or infeasible in practice \cite{ckm1}.
% Specific situations involve channels for yet-to-reach or never-to-reach locations, non-cooperative nodes, channels with large dimensions, and those operating under strict hardware or signal processing constraints .
% Additionally, CKM has shown great potential for wireless sensing, particularly in suppressing environmental noise and clutter. \cite{xuzihan} introduced a novel approach to clutter rejection enabled by CKM, leveraging a new type of CKM called the Clutter Angle Map (CLAM), which provides a promising solution for clutter suppression by incorporating environment-aware information. 

% However, the acquisition of comprehensive data measurements required for CKM construction is pretty costly and limited \cite{data}.
A fundamental problem for CKM-enabled wireless systems is how to utilize the collected location-specific channel data to construct a complete CKM. In practice, the acquisition of complete data measurements for all possible locations is costly or even impossible.
% Due to the huge coverage areas and the dynamic nature of wireless environments, data that we can obtain is often imcomplete or sparse, contributing to a significant obstacle to CKM construction.
% Millimeter-wave (mmWave) communication suffers from severe noise, which poses a significant challenge in obtaining reliable channel knowledge data \cite{data}. 
% In practical measurement scenarios,  acquiring accurate and sufficient channel knowledge data is often difficult. 
% The complexity of the physical environment further exacerbates the problem, as certain areas are inaccessible due to hazardous conditions or restrictions in confidential regions. 
Consequently, the available data are often sparse and incomplete.
The authors in \cite{xuxiaoli} provided analytical insights into the amount of data required for CKM construction and channel prediction.
% The authors in \cite{xuxiaoli} provided analytical insights into the amount of data required for accurate CKM construction and channel prediction, offering guidance on the necessary data density for both offline CKM construction and online channel prediction tasks.
% A CKM construction method based on the expectation maximization (EM) algorithm was proposed by leveraging available measurement data in conjunction with expert knowledge derived from well-established statistical channel models \cite{likun}. 
A CKM construction method based on the expectation maximization (EM) algorithm was proposed, combining available measurement data with expert knowledge from established statistical channel models \cite{likun}.
Besides, another straightforward approach is interpolation-based CKM construction, which includes K-nearest neighbors (KNN) \cite{knn}, Kriging \cite{Kriging}, and radial basis function (RBF) \cite{RBF} interpolation.
However, interpolation-based CKM construction presents inherent limitations, since sparse or unevenly distributed data points usually cause significant estimation errors. 
Furthermore, interpolation cannot inherently capture the complex physical phenomena driving channel variations, thereby restricting its ability to model intricate environments accurately.
% However, interpolation-based CKM construction has limitations, as sparse or uneven data often cause significant estimation errors. 
% Additionally, interpolation cannot capture the complex physical phenomena driving channel variations, limiting its ability to model intricate environments accurately.

% The reason is that these methods cannot learn the structure of the environment.
% Therefore, how to effectively build the complete CKM remains to be studied.

% Generative artificial intelligence (AI) has revolutionized the landscape of image generation, capable of creating high-fidelity images indistinguishable with real-world photos.
%There have been some researchers considering how to use generative AI to assist communication.
The advancement of artificial intelligence (AI) brings new opportunities for CKM construction. 
The authors in \cite{radiounet} proposed a RadioUNet architecture which maps an image representing the buildings and transmitters to a CKM image. 
The authors in \cite{gan} applied a conditional generative adversarial network (cGAN) to estimate fine-resolution CKMs from sparse channel knowledge data. 
\cite{radio_diff} utilized diffusion model to achieve sampling-free CKM construction with BS and environmental information as prompts. 
% Among a variety of generative models, diffusion models have attracted significant attention due to the ability to produce high-quality images \cite{ddpm}.
% In \cite{ddpm_channel}, authors proposed a diffusion model based channel sampling approach for fastly generating channel realizations from limited data.
Nevertheless, all of the above methods mainly rely on physical environment data and/or transmitter location information to reconstruct CKM. 
Different from the above research works, we consider the problem to directly use partially observed channel knowledge data to reconstruct the complete CKM without the use of any auxiliary environmental or positional data. 
This approach is particularly significant as it mirrors the challenges of ill-posed inverse problems frequently encountered in computer vision.

Generative AI provides a transformative framework for tackling such problems, especially diffusion model \cite{ddpm1}, offering the capability to learn the underlying implicit data distributions. 
% By leveraging these capabilities, our proposed method achieves more accurate and robust reconstructions of missing content. 
% This stands in stark contrast to traditional methods, which are often constrained by their dependency on predefined physical models or localized data, and thus struggle to capture complex global patterns. 
This is different from traditional methods that rely on predefined physical models or localized data, making it hard for them to understand complicated global patterns.
Hence, we view the incomplete CKM images as conditions to train a conditional denoiser in decoupled diffusion model (DDM) architecture proposed in \cite{ddm}.
% Simulation results show that for the two datasets CKMImageNet \cite{CKMImageNet} and RadioMapSeer \cite{radiomapseer}, our method is much better than interpolation methods in terms of mean squared error (MSE) and other indicators on the inpainted CKM images.
Simulation results show that for the CKMImageNet dataset \cite{CKMImageNet}, our method is much better than interpolation methods in terms of mean squared error (MSE) and other indicators on the inpainted CKM images.  
% In this paper, we apply generative diffusion models to recover a complete CKM from incomplete channel knowledge map, which is either partially missing or available only in a sparse, discrete format.
% By viewing the each CKM as an image, these above two scenarios can be respectively transformed as image inpainting and image super-resolution tasks.
% Our work specifically addresses CKM images restoration problem by utilizing the capabilities of generative diffusion models.
% Simulation results have shown that our method has reached the initial certification.
% Diffusion models have demonstrated predominant performance in generating high-quality images from random noise, making them ideal candidates for the task of image restoration.
% In the context of CKM reconstruction, we propose a novel application of diffusion models to handle channel knowledge data shortages.

\section{System Model}
% \subsection{Problem Formulation}
% CKM can be viewed as a transformation ${\cal M}$ from a location vector ${\bf{q}} \in {\mathbb{R}^D}$ to a channel knowledge vector ${\bf{x}} \in {\mathbb{C}^J}$ in \cite{ckm1}.
% Here, ${\bf{q}} \in {\mathbb{R}^D}$ represents a vector containing the location or virtual location of transmitters and/or receivers in wireless links, with the dimension ${D}$ varying across different application scenarios.
% The channel knowledge ${\bf{x}} \in {\mathbb{C}^J}$ denotes the pertinent channel information, where $J$ depends on the type of channel knowledge and system configuration. This transformation can be expressed as:
% ${\cal M}:{\mathbb{R}^D} \to {\mathbb{C}^J}$. 
% In a deployment consisting of $N \times N$ receivers, we define a vector ${\bf{q}} \in {\mathbb{C}^{{N^2} \times 1}}$, which indicates the physical location or virtual location of receivers. 
% We introduce a channel knowledge map ${\cal M}$ that establishes a mapping from ${\bf{q}}$ to another vector ${\bf{x}} \in {\mathbb{C}^{{N^2} \times 1}}$, where each element in ${\bf{x}}$ corresponds to the expected channel knowledge associated with the respective positions of the receivers \cite{ckm1}.

% Due to serious noise or geographical limitations, we often obtain incomplete CKM for specific geographic area. 
We consider the problem of constructing a complete CKM based on partially observed channel knowledge data for an area of interest.
% We assume that the required CKM comprises $l \times w$ locations, each carrying $D$-dimensional channel information.
% We discretize the area into $l \times w$ locations, each involving $D$ types of channel knowledge like channel gain map, angle of arrival (AoA) map, and angle of departure (AoD) map.
We discretize the area into $l \times w$ locations, each involving several types of channel knowledge like channel gain map, angle of arrival (AoA) map, and angle of departure (AoD) map.
% Due to serious noise or geographical limitations, we often only obtain the incomplete CKM. This limitation can be represented as follows:
For each channel knowledge type, the incomplete CKM can be represented as
\begin{equation}
	% {\bf{y}} = {\bf{Hx}} + {\bf{z}}
    {\bf{y}} = {\bf{Hx}},
    % {{\bf{y}}_d} = {\bf{H}}{{\bf{x}}_d},d = 1, \cdots ,D,
\end{equation}
where ${\bf{y}} \in {\mathbb{C}^{{lw} \times 1}}$ and ${\bf{x}} \in {\mathbb{C}^{{lw} \times 1}}$ denote the partially observed CKM and complete CKM, respectively.
%${\bf{H}} \in {\mathbb{C}^{{lw} \times {lw}}}$ denotes a diagonal matrix, where the diagonal elements take values of either 0 or 1 and determine which parts of ${\bf{y}}$ are missing in comparison to ${\bf{x}}$. 
${\bf{H}} \in {\mathbb{C}^{{lw} \times {lw}}}$ is a diagonal matrix representing the image mask, where ${{\bf{H}}_{ii}}$ is either $1$ and $0$, based on whether the channel knowledge at this location is observed or not.
We aim to reconstruct the complete CKM, denoted as ${\bf{\hat x}}$, from its partially observed counterpart ${\bf{y}}$.
This is a classical linear inverse problem. 
A straightforward approach to address this problem is maximizing the posterior distribution $q\left( {{\bf{x}}|{\bf{y}}} \right)$, where $q\left( {{\bf{x}}|{\bf{y}}} \right) \propto q\left( {{\bf{y}}|{\bf{x}}} \right)q\left( {\bf{x}} \right)$.
However, in scenarios where the prior distribution $q\left( {\bf{x}} \right)$ is unknown, the problem becomes underdetermined due to the rank-deficiency of the mask matrix ${\bf{H}}$.
This rank-deficiency results in the existence of multiple plausible solutions that satisfy the given observations, making the reconstruction task inherently ill-posed.
One naive method is to compute the pseudo-inverse of ${\bf{H}}$, which gives the least-squares solution as ${\bf{\hat x}} = {{\bf{H}}^\dag }{\bf{y}}$, where ${{\bf{H}}^\dag }$ denotes the pseudo-inverse of ${\bf{H}}$.
% \begin{equation}
%     {{{\bf{\tilde x}}}_d} = {{\bf{H}}^\dag }{{\bf{y}}_d}
% \end{equation}
% where ${{\bf{H}}^\dag } = {\left( {{{\bf{H}}^T}{\bf{H}}} \right)^{ - 1}}{{\bf{H}}^T}$.
However, this approach usually gives poor performance.
Besides, interpolation techniques can also be applied to complete partially missing data, such as linear, bilinear and Kriging interpolation. 
But various interpolation methods may result in low-quality outcomes, as they often estimate missing data values solely based on the known values surrounding them.

\begin{figure}[t]
	\centerline{\includegraphics[width=8.5cm]{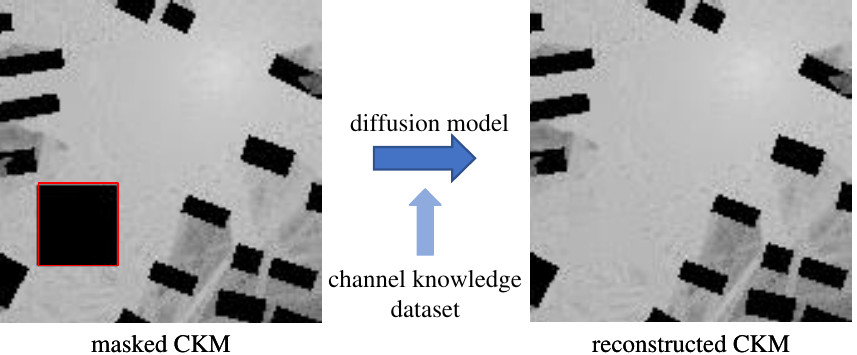}}
	\caption{CKM image inpainting task framework.}
	\label{problem}
\end{figure}

% Consequently, the key idea of the proposed approach is to learn $q\left( {\bf{x}} \right)$, which aligns well with the generative diffusion model's thought. 
% By leveraging a generative model based on a dataset $\left\{ {{\bf{x}}} \right\}$, we can better capture the complex relationships and structures inherent in the data, thus achieving CKM inpainting task as plotted in Fig. \ref{problem}.
The core concept of the proposed approach focuses on effectively learning the prior distribution $q\left( {\bf{x}} \right)$, which aligns seamlessly with the principles of generative diffusion models in \cite{ddpm1}.
By leveraging diffusion model trained on a dataset $\left\{ {{\bf{x}}} \right\}$, the method captures the intricate relationships and underlying structures within the data. 
This enables the reconstruction of missing information in CKMs, as illustrated in Fig.~\ref{problem}.
% For simplicity, we let ${\bf{x}}$ replace ${\bf{x}}$, representing any type of CKMs.

\begin{figure*}[t]
	\centerline{\includegraphics[width=12.5cm]{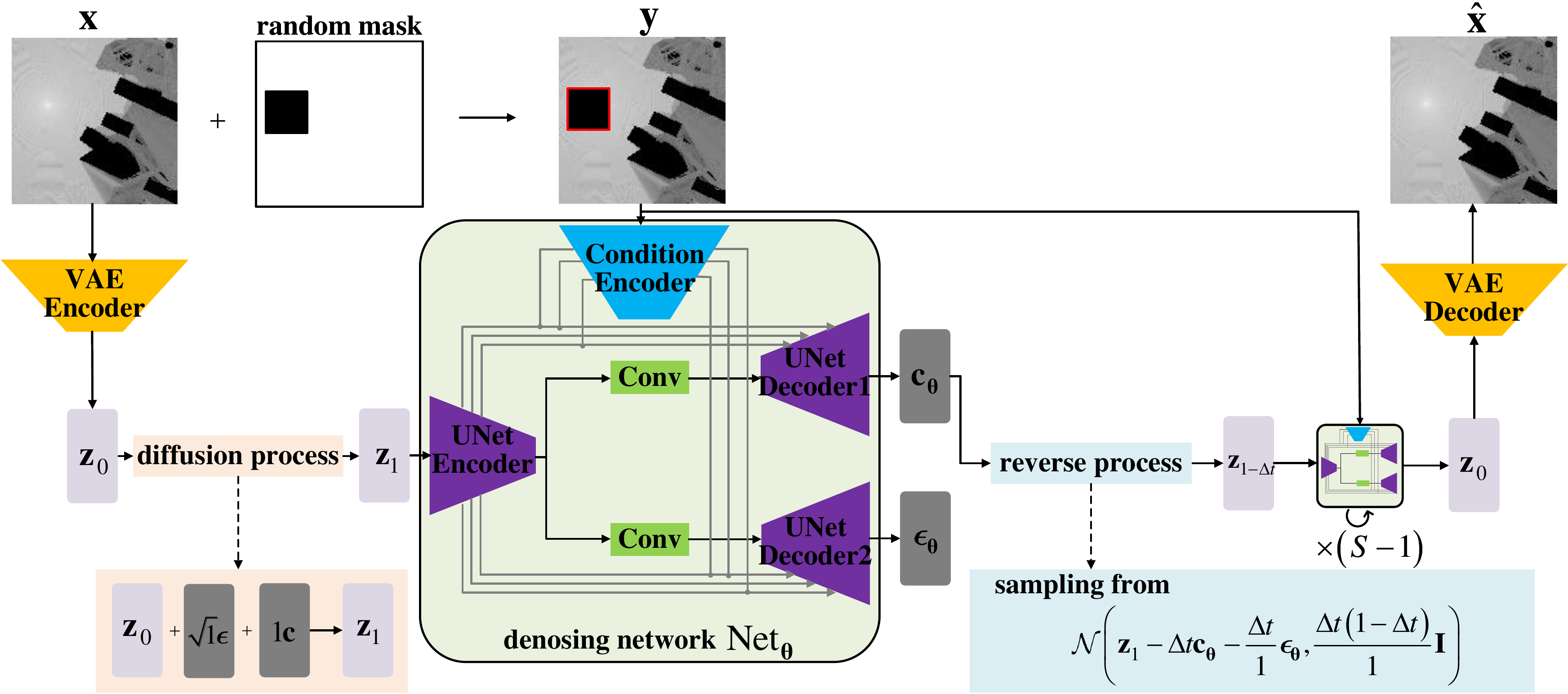}}
	\caption{Conditional decoupled diffusion model for generative CKM construction.}
	\label{ddm}
\end{figure*}

\section{CKM Construction via Diffusion Models}
\subsection{Principles of Diffusion Models}
% Traditional interpolation methods such as KNN to do the reconstruction of images usually use the feature information of a single image to complement, and the effect is not ideal. 
% Therefore, in this section we introduce the principle of generated diffusion model, which is further applied to construct CKM images. 

% We propose to use conditional diffusion models to solve both the inpainting and super-resolution tasks on CKM images construction.
% Diffusion model is one of the most popular generative models that uses the concept of diffusion process to generate new data samples.
% Diffusion model is effectively used to learn complex data distributions and then generating high-quality samples, such as images, text, and audio.
% The core of denoising diffusion probabilistic models (DDPM) proposed in \cite{ddpm1} is a probabilistic process that gradually distorts the input data by adding random Gaussian noise over many time steps, until it achieves a near Gaussian distribution.
% Diffusion model is effectively used to learn complex data distributions and then generating high-quality samples, such as images, text, and audio.
% The core of denoising diffusion probabilistic models (DDPM) proposed in \cite{ddpm1} is to maximize the log-likelihood function of $q\left( {\bf{x}} \right)$.
The core principle of denoising diffusion probabilistic models (DDPM), proposed in \cite{ddpm1}, is to model the data distribution $q\left( {\bf{x}} \right)$ by minimizing a variational bound on the negative log-likelihood. 
This framework utilizes a forward process to gradually corrupt the data by adding Gaussian noise over a sequence of time steps $t \in \left[ {1,T} \right]$, and a reverse process to reconstruct the original data by denoising step-by-step.
% The forward diffusion process gradually destroys the data adding additive Gaussian noise over a sequence of time steps $t \in \left[ {1,T} \right]$.
Starting from an initial data ${{\bf{x}}_0}$, the forward process is defined as:
\begin{equation}
	q\left( {{{\bf{x}}_t}|{{\bf{x}}_{t - 1}}} \right) = {\cal N}\left( {{{\bf{x}}_t};\sqrt {1 - {\beta _t}} {{\bf{x}}_{t - 1}},{\beta _t}{\bf{I}}} \right),
	\label{forward process}
\end{equation}
where ${\beta _t}$ is a variance schedule that determines how much noise is added at each step $t$, and ${\bf{I}}$ is the identity matrix.
The process is continued until time step $T$, when the original data are completely corrupted by noise.
It is known that the forward process admits sampling ${{\bf{x}}_t}$ at an arbitrary timestep $t$ in closed form:
\begin{equation}
	q\left( {{{\bf{x}}_t}|{{\bf{x}}_0}} \right) = {\cal N}\left( {{{\bf{x}}_t};\sqrt {{{\bar \alpha }_t}} {{\bf{x}}_0},\left( {1 - {{\bar \alpha }_t}} \right){\bf{I}}} \right),
	\label{forward x0}
\end{equation}
where ${{\bar \alpha }_t} = \prod\nolimits_{s = 1}^t {{\alpha _s}} $ and ${{\alpha _t} = 1 - {\beta _t}}$. 

% \textbf{Reverse diffusion process}:
The reverse process is opposite of the forward diffusion process, which aims to recover the original data by eliminating  the added noise gradually.
% By defining a Markov chain that starts from the heavily-noised data ${\bf{x}_T}$ and moves backward through time towards ${x_0}$:
By defining a Markov chain, it starts with the highly noisy data ${{\bf{x}}_T}$ and moves in reverse to ${{\bf{x}}_0}$. The specific process from ${{\bf{x}}_t}$ to ${{\bf{x}}_{t-1}}$ can be expressed by
\begin{equation}
	{p_{\bm{\theta}}}\left( {{{\bf{x}}_{t - 1}}|{{\bf{x}}_t}} \right) = {\cal N}\left( {{{\bf{x}}_{t - 1}};{{\bm{\mu}}_{\bm{\theta}} }\left( {{{\bf{x}}_t},t} \right),{{\bm{\Sigma }}_{\bm{\theta}} }\left( {{{\bf{x}}_t},t} \right)} \right).
	\label{reverse process}
\end{equation}
Here, ${{\bm{\mu} _{\bm{\theta}} }}$ and ${{\bm{\Sigma} _{\bm{\theta}} }}$ are learned functions parameterized by ${\bm{\theta}} $.
Specifically, ${{\bm{\mu} _{\bm{\theta}} }}$ predicts the mean of the Gaussian distribution and ${{\bm{\Sigma} _{\bm{\theta}} }}$ defines the covariance matrix.

% \textbf{Loss function}:
The goal of training a diffusion model is to minimize the distance between the real data distribution and the generated data distribution.
It has been shown that minimizing the variational bound on the negative log likelihood can be simplified to minimizing \cite{ddpm1}
\begin{equation}
	% {\cal L} =  - {\mathbb{E}_{{{\bf{x}}_0}}}\left[ {\log {p_\theta }\left( {{{\bf{x}}_0}} \right)} \right]\mathop  \approx \limits^{\left( a \right)} {\mathbb{E}_{t,{{\bf{x}}_0},{\bm{\epsilon }}}}\left[ {{{\left\| {{\bm{\epsilon }} - {{\bm{\epsilon }}_\theta }\left( {{{\bf{x}}_t},t} \right)} \right\|}^2}} \right]
    {\cal L} =  {\mathbb{E}_{t,{{\bf{x}}_0},{\bm{\epsilon }}}}\left[ {{{\left\| {{\bm{\epsilon }} - {{\bm{\epsilon }}_{\bm{\theta}} }\left( {{{\bf{x}}_t},t} \right)} \right\|}^2}} \right],
	\label{loss function}
\end{equation}
where $\bm{\epsilon}  \sim {\cal N}\left( {{\bf{0,I}}} \right)$ is sampled noise added at each forward diffusion step, and ${{\bf{x}}_t} = \sqrt {{{\bar \alpha }_t}} {{\bf{x}}_0} + \sqrt {1 - {{\bar \alpha }_t}} {\bm{\epsilon }}$.
A UNet structured neural network ${{{\bm{\epsilon }}_{\bm{\theta}} }\left( {{{\bf{x}}_t},t} \right)}$ is applied to predict the noise.
% Hence, the core of diffusion model training is to predict noise added in the forward process. 
% $\bm{\epsilon}  \sim {\cal N}\left( {{\bf{0,I}}} \right)$ is sampled noise, and ${\bm{\epsilon} _\theta }$ is the noise prediction network that needs to be trained.

% \textbf{U-Net network to predict noise}:
% A U-Net structured neural network (depicted in Fig.\ref{unet}) is applied to predict the noise ${{{\bm{\epsilon }}_\theta }\left( {{{\bf{x}}_t},t} \right)}$ which is obtained from a corrupted input and related time step index.
% During training stage, the network is to minimize the mean square deviation between its noise predictions and the actual noise added.
% The U-Net's design is characterized by a dual pathway architecture: a downsampling stage and an upsampling stage.
% The downsampling process employs a series of convolutional operations with max-pooling stages, which enables the extraction of gradually more abstract feature representations.
% On the other hand, the upsampling process utilizes a series of transposed convolutions to reconstruct the spatial dimensions of the input, gradually restoring data details.

% \begin{figure}[t]
% 	\centerline{\includegraphics[width=8.5cm]{unet.png}}
% 	\caption{The U-Net architecture used for denoising.}
% 	\label{unet}
% \end{figure}

\subsection{Conditional Diffusion Model for CKM Inpainting}

% \begin{figure*}[t]
% 	\centerline{\includegraphics[width=15.5cm]{con_ddm.png}}
% 	\caption{Conditional Diffusion Model network architecture.}
% 	\label{ddm}
% \end{figure*}

In order to further improve the quality and speed of CKM reconstruction, we apply conditional DDM proposed in \cite{ddm} to solve this inverse problem. 
Different from DDPM, DDM converts the normal image-to-noise forward diffusion process into two stages:(i) attenuate the image component by an image-to-zero mapping, (ii) increase the noise component by a zero-to-noise mapping.
By utilizing decoupled diffusion strategy, the forward diffusion process is represented by
\begin{equation}
	{{\bf{x}}_t} = {{\bf{x}}_0} + \int_0^t {{{\bf{f}}_t}dt}  + \int_0^t {d{{\bf{w}}_t}},
	\label{forward1 ddm}
\end{equation}
where ${{\bf{x}}_0} + \int_0^t {{{\bf{f}}_t}dt}$ denotes the image attenuation process and $\int_0^t {d{{\bf{w}}_t}}$ represents the noise enhancement process. 
${{\bf{w}}_t}$ is the standard Wiener process, and ${{{\bf{f}}_t}}$ is a differentiable function of ${t}$ that allows arbitrary sampling steps, thus accelerating the sampling process. 
In this paper, we set ${{{\bf{f}}_t}}$ as a simple but effective function ${{{\bf{f}}_t} = \bf{c}}$.
Note that $t$ takes the range $\left[ {0,1} \right]$, different from DDPM. 
To achieve the forward diffusion process from ${{\bf{x}}_0}$ to noise, we need to ensure that the transformation effectively transitions the initial state into a noise-dominated state.
Thus, ${{\bf{x}}_0} $ follows the distribution $q\left( {{{\bf{x}}_0}} \right)$ and ${{\bf{x}}_0} + \int_0^1 {{{\bf{f}}_t}dt}  = {\bf{0}}$, satisfying that ${{\bf{x}}_1}$ is distributed as ${\cal N}\left( {{\bf{0}},{\bf{I}}} \right)$. 
In other words, ${\bf{c}} =  - {{\bf{x}}_0}$.

Concisely, the forward process can be rewritten as:
\begin{equation}
	% q\left( {{{\bf{x}}_t}|{{\bf{x}}_0}} \right) = {\cal N}\left( {{{\bf{x}}_t};{{\bf{x}}_0} + \int_0^t {{{\bf{f}}_t}dt} ,t{\bf{I}}} \right).
    q\left( {{{\bf{x}}_t}|{{\bf{x}}_0}} \right) = {\cal N}\left( {{{\bf{x}}_t};{{\bf{x}}_0} + \int_0^t {{\bf{c}}dt} ,t{\bf{I}}} \right).
	\label{forward ddm}
\end{equation}
Therefore, we can sample ${{\bf{x}}_t}$ by ${{\bf{x}}_t} = \left( {1 - t} \right){{\bf{x}}_0} + \sqrt t {\bm{\epsilon }}$, where ${\bm{\epsilon }} \sim {\cal N}\left( {{\bf{0}},{\bf{I}}} \right)$. 

Unlike DDPM which uses discrete time Markov chain, the reverse process in DDM employs continuous-time Markov chain with the smallest time step ${\Delta t \to {0^ + }}$ and we use conditional distribution ${q\left( {{{\bf{x}}_{t - \Delta t}}|{{\bf{x}}_t},{{\bf{x}}_0}} \right)}$ to approximate ${q\left( {{{\bf{x}}_{t - \Delta t}}|{{\bf{x}}_t}} \right)}$:
\begin{equation}
\small
	% q\left( {{{\bf{x}}_{t - \Delta t}}|{{\bf{x}}_t},{{\bf{x}}_0}} \right) = {\cal N}\left( {{{\bf{x}}_t} + \int_t^{t - \Delta t} {{{\bf{f}}_t}dt}  - \frac{{\Delta t}}{{\sqrt t }}{\bm{\epsilon }},\frac{{\Delta t\left( {t - \Delta t} \right)}}{t}{\bf{I}}} \right).
	q\left( {{{\bf{x}}_{t - \Delta t}}|{{\bf{x}}_t},{{\bf{x}}_0}} \right) = {\cal N}\left( {{{\bf{x}}_t} - \Delta t{\bf{c}} - \frac{{\Delta t}}{{\sqrt t }}{\bm{\epsilon }},\frac{{\Delta t\left( {t - \Delta t} \right)}}{t}{\bf{I}}} \right).
    \label{reverse ddm}
\end{equation}

\begin{figure*}
    \begin{equation}
        {p_{\bm{\theta }}}\left( {{{\bf{x}}_{t - \Delta t}}|{{\bf{x}}_t},e\left( {\bf{y}} \right)} \right) = {\cal N}\left( {{{\bf{x}}_{t - \Delta t}};{{\bf{x}}_t} - {\Delta t{{\bf{c}}_{\bm{\theta }}}\left( {{{\bf{x}}_t},t|e\left( {\bf{y}} \right)} \right)}  - \frac{{\Delta t}}{{\sqrt t }}{{\bm{\epsilon }}_{\bm{\theta }} }\left( {{{\bf{x}}_t},t|e\left( {\bf{y}} \right)} \right),\frac{{\Delta t\left( {t - \Delta t} \right)}}{t}{\bf{I}}} \right).
        \label{reverse_ddm_con}
    \end{equation}
    \end{figure*}
% From the reverse process in \eqref{reverse ddm}, it is known that we need to predict signal attenuation term ${{\int_t^{t - \Delta t} {{{\bf{f}}_t}dt} }}$ and noise term ${{\bm{\epsilon }}}$ to restore orignal image ${{{{\bf{x}}_0}}}$.
From the reverse process in \eqref{reverse ddm}, signal attenuation term $- \Delta t{\bf{c}}$ and noise term ${{\bm{\epsilon }}}$ are unknown. 
% Hence, we use ${p_{\bm{\theta }}}\left( {{{\bf{x}}_{t - \Delta t}}|{{\bf{x}}_t}} \right)$ to approximate $q\left( {{{\bf{x}}_{t - \Delta t}}|{{\bf{x}}_t},{{\bf{x}}_0}} \right)$ by using a neural network, where the UNet architecture with two decoder branches is utilized to predict simultaneously ${{\int_t^{t - \Delta t} {{{\bf{f}}_t}dt} }}$ and ${{\bm{\epsilon }}}$.
% Hence, we use ${p_{\bm{\theta }}}\left( {{{\bf{x}}_{t - \Delta t}}|{{\bf{x}}_t}} \right)$ to approximate $q\left( {{{\bf{x}}_{t - \Delta t}}|{{\bf{x}}_t},{{\bf{x}}_0}} \right)$ and by using a neural network, where the UNet architecture with two decoder branches is utilized to predict simultaneously ${\bf{c}}$ and ${{\bm{\epsilon }}}$.
Hence, we use ${p_{\bm{\theta }}}\left( {{{\bf{x}}_{t - \Delta t}}|{{\bf{x}}_t}} \right)$ to approximate $q\left( {{{\bf{x}}_{t - \Delta t}}|{{\bf{x}}_t},{{\bf{x}}_0}} \right)$ and predict simultaneously ${\bf{c}}$ and ${{\bm{\epsilon}}}$ by utilizing a modified UNet architecture ${\rm{Net}}_{\bm{\theta}}$, where two stacked convolutional layers are added to create two UNet decoder branches.
In addition, an encoder is employed to extract multi-level features of masked image ${\bf{y}}$ as $e\left( {\bf{y}} \right)$ for a conditional DDM to inpaint CKM images.
Therefore, $ {p_{\bm{\theta }}}\left( {{{\bf{x}}_{t - \Delta t}}|{{\bf{x}}_t},e\left( {\bf{y}} \right)} \right)$ can be expressed as \eqref{reverse_ddm_con}.
% \begin{figure*}
% \begin{equation}
% 	{p_{\bm{\theta }}}\left( {{{\bf{x}}_{t - \Delta t}}|{{\bf{x}}_t},e\left( {\bf{y}} \right)} \right) = {\cal N}\left( {{{\bf{x}}_{t - \Delta t}};{{\bf{x}}_t} - {\Delta t{{\bf{c}}_{\bm{\theta }}}\left( {{{\bf{x}}_t},t|e\left( {\bf{y}} \right)} \right)}  - \frac{{\Delta t}}{{\sqrt t }}{{\bm{\epsilon }}_{\bm{\theta }} }\left( {{{\bf{x}}_t},t|e\left( {\bf{y}} \right)} \right),\frac{{\Delta t\left( {t - \Delta t} \right)}}{t}{\bf{I}}} \right).
% 	\label{reverse_ddm_con}
% \end{equation}
% \end{figure*}
% ${{\bf{f}}_t}$ is an analytic function with respect to $t$, which is determined by its parameters $\bm{\phi}$.  
% Therefore, we utilize the U-Net architecture with two decoder branches which is plotted in Fig.\ref{ddm} to predict $\bf{\phi}$ and $\bf{\epsilon}$.
Accordingly, the training objective is
\begin{equation}
	% \mathop {\min }\limits_\theta  {\mathbb{E}_{q\left( {{{\bf{x}}_0}} \right)}}{\mathbb{E}_{q\left( {\bm{\epsilon }} \right)}}\left[ {{{\left\| { {\bm{\phi}}- {{\bm{\phi}}_\theta }} \right\|}^2} + {{\left\| { {\bm{\epsilon }}- {{\bm{\epsilon }}_\theta }} \right\|}^2}} \right]
	\mathop {\min }\limits_{\bm{\theta}}  {\mathbb{E}_{t,{{\bf{x}}_0},{\bm{\epsilon }},{\bf{y}}}}\left[ {{{\left\| {{\bf{c}} - {{\bf{c}}_{\bm{\theta}} }\left( {{{\bf{x}}_t},t|e\left( {\bf{y}} \right)} \right)} \right\|}^2} + {{\left\| {{\bm{\epsilon }} - {{\bm{\epsilon }}_{\bm{\theta}} }\left( {{{\bf{x}}_t},t|e\left( {\bf{y}} \right)} \right)} \right\|}^2}} \right].
    \label{ddm loss}
\end{equation}

To sum up, the specific training and sampling procedures for CKM construction are summarized in Algorithm~\ref{train_algorithm} and Algorithm~\ref{sample_algorithm}, respectively.
The architecture of the network is demonstrated in Fig.~\ref{ddm}. 
% Like latent diffusion models \cite{ldm}, a variational autoencoder (VAE) is employed to map the real pixel space ${\bf{x}}$ to a latent space ${\bf{z}}$, which aims to reduce computational requirements and training time.
A variational autoencoder (VAE) is employed to map the pixel space ${\bf{x}}$ to the latent space ${\bf{z}}$, aiming to reduce computational requirements and training time.
% Therefore, we utilize the U-Net architecture with two decoder branches which is demonstrated in Fig.\ref{ddm} to predict simultaneously $\bm{\phi}$ and $\bm{\epsilon}$.
% Besides, like latent diffusion models \cite{ldm}, a Variational Autoencoder (VAE) is employed to map the real pixel space to a latent space, which aims to reduce computational requirements and training time.
% And for inpainting CKM images, we use a conditional DDM architecture. 
% Besides, we apply a Swin-B encoder \cite{swin} to extract multi-level features of the conditioned input, and concatenate these features and the image features with the same levels as the UNet decoder's inputs.
Besides, we apply a Swin-B encoder \cite{swin} to extract multi-level features of the conditioned input, and concatenate these features with the image features at the UNet decoder's input levels.

\begin{algorithm}[t]
\caption{The training algorithm for CKM construction}
 \begin{algorithmic}[1]
\State \textbf{Initialize}: $i = 0$, number of iterations: $N$, network parameters: ${\bm{\theta }}$
\While{$i < N$}
    \State $t \sim \text{Uniform}(0, 1)$, ${\bf{x}}_0 \sim q({\bf{x}}_0)$, ${\bf{c}}= - {{\bf{x}}_0}$, ${\bf{y}} = {\bf{H}}{\bf{x}}_0$ where ${\bf{H}}$ is a random mask matrix, $\bm{\epsilon} \sim \mathcal{N}({\bf{0}}, {\bf{I}})$
    % \Statex ${\bf{y}} = {\bf{A}}{\bf{x}}_0$, 
    \State $e({\bf{y}}) = \text{encoder}({\bf{y}})$
    \State ${\bf{x}}_t = {\bf{x}}_0 + t {\bf{c}} + \sqrt{t} \bm{\epsilon}$
    \State ${\bf{c}}_{\bm{\theta}}, \bm{\epsilon}_{\bm{\theta}} = \text{Net}_{\bm{\theta}}({\bf{x}}_t, t, e({\bf{y}}))$
    \State \textbf{Take gradient descent step on}
    
    ${\nabla_{\bm{\theta}}} \left( \left\| {\bf{c}} - {\bf{c}}_{\bm{\theta}}({\bf{x}}_t, t | e({\bf{y}})) \right\|^2 + \left\| \bm{\epsilon} - \bm{\epsilon}_{\bm{\theta}}({\bf{x}}_t, t | e({\bf{y}})) \right\|^2 \right)$
    \State $i = i + 1$
\EndWhile
\State \textbf{return} $\bm{\theta}$
\end{algorithmic}
\label{train_algorithm}
\end{algorithm}

\begin{algorithm}[t]
\caption{The sampling algorithm for CKM construction}
\begin{algorithmic}[1]
\State \textbf{Initialize}: ${\bf{x}}_1 \sim \mathcal{N}({\bf{0}}, {\bf{I}})$, ${\bf{y}}$, number of sampling steps: $S$, sampling interval: $\Delta t = 1/S$, ${\rm{Net}}_{\bm{\theta}}$
\While{$t > 0$}
    \State ${\bm{\tilde{\epsilon}}} \sim \mathcal{N}({\bf{0}}, {\bf{I}})$
    \State $e({\bf{y}}) = \text{encoder}({\bf{y}})$
    \State ${\bf{c}}_{\bm{\theta}}, {\bm{\epsilon}}_{\bm{\theta}} = \text{Net}_{\bm{\theta}}({\bf{x}}_t, t, e({\bf{y}}))$
    \State ${\bf{x}}_t = {\bf{x}}_t - \Delta t {\bf{c}}_{\bm{\theta}} - \frac{\Delta t}{\sqrt{t}} {\bm{\epsilon}}_{\bm{\theta}} + \sqrt{\frac{\Delta t(t - \Delta t)}{t}} \bm{\tilde{\epsilon}}$
    \State $t = t - \Delta t$
\EndWhile
\State \textbf{return} ${\bf{x}}_t$
\end{algorithmic}
\label{sample_algorithm}
\end{algorithm}

\section{Numerical Results}
% In order to demonstrate the advantages of diffusion model for CKM reconstruction, we evaluate two datasets, i.e. CKMImageNet dataset and RadioMapSeer dataset. 
% In the following, we first introduce specific implementation of our proposed CKMImageNet and RadioMapSeer. 
% Accordingly, we evaluate the performance of diffusion model on CKM image inpainting task, compared with several interpolation methods for CKMImageNet and RadioMapSeer datasets, respectively.
% In the following, we first introduce specific implementation of our proposed CKMImageNet dataset and experiment setup. 
% For the CKM image inpainting problem, we divide it into two scenarios according to whether the building is masked or not. 
% We evaluate the performance of diffusion model on CKM image inpainting task, compared with several interpolation methods for two scenarios, respectively. 
% In addition, to further prove the effectiveness of our algorithm for locating the BS, we evaluate the performance on two datasets, i.e., CKMImageNet dataset and RadioMapSeer dataset.

\subsection{Dataset and Simulation Setup}
The CKMImageNet dataset is used to train and verify our model \cite{CKMImageNet}. 
The CKMImageNet dataset is a mixed dataset of numerical data, images of different resolutions and environment maps. 
The data are obtained from ray-tracing by using commercial software \textit{Wireless Insite}\textsuperscript{1},
which simulates electromagnetic wave propagation with high fidelity as it models the interactions of waves with the environment, including reflections, diffractions, and scatterings. 
Specifically, six reflections and one diffraction were configured for the CKMImageNet dataset.
The simulation settings involve a transmit power of $0$ dBm and a carrier frequency of $28$ GHz, with the transmitter and receiver heights set to 10 meters and 1 meter, respectively.
% Simulation parameters are listed in table.\ref{ckm_parameter}. 
% \footnotetext[1]{\url{https://www.remcom.com/wireless-insite-em-propagation-software}}
% \footnotetext[2]{\url{https://www.openstreetmap.org/}}
The environment map involves cities like Beijing, Shanghai, London, etc, and the city maps are mainly provided by \textit{OpenStreetMap}\textsuperscript{2}. 
% Currently, the CKMImageNet dataset consists of more than 50000 images of the resolution of $ 64\times 64 $ and over 30000 images of the resolution of $ 128\times 128 $, where each pixel corresponds to a $2 \times 2$ square meters in the real-world terrain.
Currently, the CKMImageNet dataset consists of over 30000 images of the resolution of $ 128\times 128 $, where adjacent pixel points are spaced two meters apart in the real-world terrain.
% In addition, different from many other datasets for communication which contain only data of channel gain, the CKMImageNet dataset also contains other channel knowledge like AoA, AoD and time of arrival.  
% However, our proposed scheme only verifies the inpainting effect of the channel gain map.
% \footnotetext[1]{\url{https://www.remcom.com/wireless-insite-em-propagation-software}}
% \footnotetext[2]{\url{https://www.openstreetmap.org/}}
% \begin{table}[!t]
% 	\renewcommand{\arraystretch}{1.3}
% 	\caption{CKMImageNet Simulation Parameters}
% 	\centering
% 	\label{ckm_parameter}
% 	\begin{tabular}{c c}
% 		\hline\hline                                              \\[-4mm]
% 		\multicolumn{1}{c}{Parameter} & \multicolumn{1}{c}{value} \\[0.5ex] \hline
% 		Carrier frequency            & $28$GHz                     \\
% 		Bandwidth     & $1$MHz                       \\
% 		Waveform               & Sinusoid                    \\
% 		Antenna              & Isotropic                  \\
% 		Number of reflections                 & $6$                 \\
% 		Number of diffractions                  & $1$                      \\
% 		Number of scatterings        & $0$                 \\
%         X-axis minimum interval units                       & $\Delta x = 2 $m                     \\
% 		Y-axis minimum interval units                     & $\Delta y = 2 $m                     \\
% 		\hline\hline
% 	\end{tabular}
% \end{table}
% However, our proposed scheme have only verified the inpainting effect of the channel gain map.
% Next we will introduce the relationship between the images and the channel knowledge data. 
As for the channel gain images, the channel gain values are between $-50$dB and $-250$dB, and the values are linearly mapped to $[0,255]$ to form the grayscale images that the diffusion model requires. 
In addition, we assume that the values in areas where there are buildings are the minimum value of $-250$, which is map to gray scale $0$ and shows black in grayscale image. 
% For AoA and AoD maps, the way grayscale images are formed is similar. 
% The difference lies in the assumption that the angle value at a location where a signal cannot be received is set to $- {200^ \circ }$, which is not within the actual angle range of $- {180^ \circ }$ to $ {180^ \circ }$ , thus allowing us to distinguish different locations in the map for subsequent model training.

\footnotetext[1]{\url{https://www.remcom.com/wireless-insite-em-propagation-software}}
\footnotetext[2]{\url{https://www.openstreetmap.org/}}

% In our task, images of the resolution of $ 128\times 128 $ are used to finish the task.

% \subsection{Simulation Setup}
We use 20000 $ 128\times 128 $ channel gain images as training dataset, and 2000 channel gain images as testing dataset. 
We train our CKM image inpainting model in two steps. 
% First, a VAE is trained to map the real pixel space to the latent space to reduce the computational resource consumption of the diffusion model.
First, a VAE is trained to map the pixel space to the latent space.  
An AdamW optimizer is utilized with a scheduled learning rate from 5e-6 to 1e-6. Next, we train a conditional DDM using the AdamW optimizer. 
The model is trained for $300,000$ iterations with a batch size of $48$ and the smallest time step is $1 \times {10^4}$.
% The hyper-parameters of the trained DDM are listed in table. \ref{Parameter_t}. 
All the above training and sampling processes can be implemented on a NVIDIA GeForce RTX 4090.

% \begin{table}[!t]
%     \renewcommand{\arraystretch}{1.5}
%     \caption{Training hyperparameter settings}
%     \centering
%     \label{Parameter_t}
%     \begin{tabular}{c c | c c}
%         \hline\hline                                              \\[-4mm]
%         % \multicolumn{1}{l}{Parameter} & \multicolumn{1}{l}{Value} & \multicolumn{1}{l}{Value} \\[0.5ex] \hline
%         Dataset                       & CKMImageNet               & RadioMapSeer                \\
%         Image size                    & $128 \times 128$          & $256 \times 256$            \\
%         Batch size                    & $48$                      & $32$                        \\
%         Iterations                    & $400000$                  & $300000$                    \\
%         Learning rate                 & $4 \times {10^{ - 5}} \sim 4 \times {10^{ - 6}}$  &  $4 \times {10^{ - 5}} \sim 4 \times {10^{ - 6}}$  \\
%         Feature channels              & $96$                      & $96$                        \\
%         Number of blocks              & $2$                       & $2$                          \\
%         Channel multiplier            & $[1,2,4,8]$               & $[1,2,4,8]$                   \\
% 		Smallest time step            & $1 \times {10^{ - 4}}$    & $1 \times {10^{ - 4}}$         \\
%         \hline\hline
%     \end{tabular}
% \end{table}

\subsection{Simulation Results}

% According to our observation, the performance of CKM image inpainting is very sensitive to building information. 
% Because in this case it is insufficient for the model to just take into account the laws of electromagnetic propagation. 
% Building areas often lack actual signal measurements and are represented as solid black, rather than with real gain, making it challenging for the model to capture true channel characteristics. 
% Due to the lack of signal measurements at the locations of the buildings, there is no real gain values.
We complete CKM images for two scenarios: mask covers the buildings and mask does not cover the buildings.
% It is known that channel knowledge cannot be measured at the location obscured by buildings. 
% The error between pixels where buildings are located in ground truth image and corresponding ones in reconstructed image is not considered. 
Besides, to better illustrate the effectiveness of CKM reconstruction, we convert pixel values in gray images into CKM ground truth values and further calculate error metrics for all locations without building occupancy, i.e., MSE, normalized MSE (NMSE), root MSE (RMSE) and mean absolute error (MAE).

\begin{figure*}[htbp]  % 使用 figure* 环境以跨越两栏
    \centering
    \begin{minipage}{0.125\textwidth}  % 每个子图宽度
        \centering
        \includegraphics[width=\linewidth]{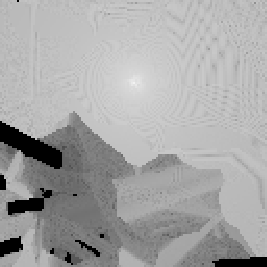}  % 替换为你的图片路径
        \subcaption{ground truth}
    \end{minipage}
    \hfill  % 水平填充以分隔子图
    \begin{minipage}{0.125\textwidth}
        \centering
        \includegraphics[width=\linewidth]{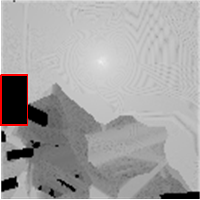}
        \subcaption{inpaint}
    \end{minipage}
    \hfill
    \begin{minipage}{0.125\textwidth}
        \centering
        \includegraphics[width=\linewidth]{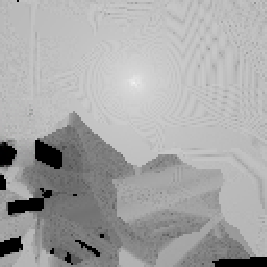}
        \subcaption{\textbf{proposed}}
    \end{minipage}
    \hfill
    \begin{minipage}{0.125\textwidth}
        \centering
        \includegraphics[width=\linewidth]{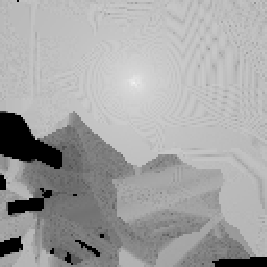}
        \subcaption*{KNN}
    \end{minipage}
    \hfill
    \begin{minipage}{0.125\textwidth}
        \centering
        \includegraphics[width=\linewidth]{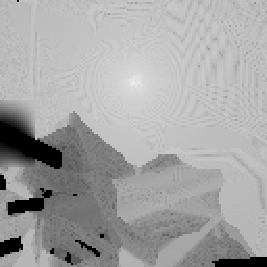}
        \subcaption{Kriging}
    \end{minipage}
    \hfill
    \begin{minipage}{0.125\textwidth}
        \centering
        \includegraphics[width=\linewidth]{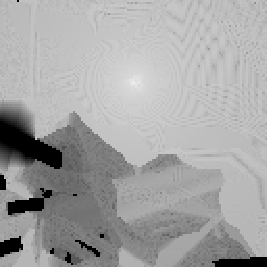}
        \subcaption{Bilinear}
    \end{minipage}
    \hfill
    \begin{minipage}{0.125\textwidth}
        \centering
        \includegraphics[width=\linewidth]{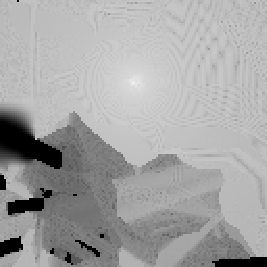}
        \subcaption{RBF}
    \end{minipage}
    
    % \caption{Visualization results of different methods for CKMImageNet dataset when mask convering buildings. (a) Ground truth. (b) Masked image. (c) Diffusion model. (d) KNN. (e) Kriging. (f) Bilinear. (g) RBF. }
    \caption{Visualization results of different methods when buildings are masked.}
    \label{fig:ckm_seven_images_buliding}
\end{figure*}

Fig.~\ref{fig:ckm_seven_images_buliding} illustrates the inpainting effect of the scenario where the mask covers the buildings. 
Here we compare our method with several interpolation methods, such as KNN, Kriging, Bilinear \cite{bilinear} and RBF interpolation.
% It is known that channel knowledge cannot be measured at the location obscured by buildings. 
% Hence, the error between pixels where buildings are located in ground truth image and corresponding ones in reconstructed image is not considered. 
% In addition, to better illustrate the effectiveness of our restoration, we convert pixel values in gray images into CKM truth values and further calculate error metrics, i.e., MSE, Normalized MSE (NMSE), RMSE and Mean Absolute Error (MAE).
Table~\ref{ErrorMetricsTable1} plots the average error metrics of inpainting $1000$ CKM images when masking buildings with the five methods. 
It is observed that our proposed diffusion based model significantly outperforms significantly the benchmark approaches. 
% However, the performance is still need to be further improved in this situation. 

\begin{table}[!t]
    \renewcommand{\arraystretch}{1.5}  % 增加行间距
    \caption{Performance Comparison With Masking Buildings}
    \centering
    \label{ErrorMetricsTable1}
    \setlength{\tabcolsep}{4pt}
    \begin{tabular}{c|c|c|c|c|c}
        \hline
        % \multicolumn{1}{c|}{} & \multicolumn{5}{c|}{\textbf{CKMImageNet}} & \multicolumn{5}{c}{\textbf{RadioMapSeer}} \\ \hline
        % \diagbox{\textbf{Metric}}{\textbf{Method}} & \textbf{KNN} & \textbf{Kriging} & \textbf{Bilinear} & \textbf{RBF} & \textbf{Diff} \\ \hline
        \multicolumn{1}{c|}{} & \textbf{KNN} & \textbf{Kriging} & \textbf{Bilinear} & \textbf{RBF} & \textbf{Proposed} \\ \hline
        \textbf{MSE$({\rm{d}}{{\rm{B}}^2})$}  & $4359.266$ & $1868.018$ & $3526.219$ & $2910.859$ & \textbf{427.299} \\ \hline
        \textbf{NMSE}  & $0.0722$ & $0.0309$ & $0.0584$ & $0.0482$ & \textbf{0.0071}  \\ \hline
        \textbf{RMSE$({\rm{dB}})$}  & $66.0247$ & $43.2206$ & $59.3820$ & $53.9524$ & \textbf{20.6712}  \\ \hline
        \textbf{MAE$({\rm{dB}})$}   & $36.7986$ & $29.8942$ & $40.2318$ & $34.0901$ & \textbf{9.9813} \\ \hline
    \end{tabular}
\end{table}

\begin{figure*}[htbp]  % 使用 figure* 环境以跨越两栏
    \centering
    \begin{minipage}{0.125\textwidth}  % 每个子图宽度
        \centering
        \includegraphics[width=\linewidth]{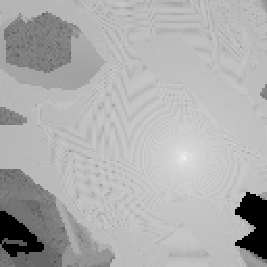}  % 替换为你的图片路径
        \subcaption{ground truth}
    \end{minipage}
    \hfill  % 水平填充以分隔子图
    \begin{minipage}{0.125\textwidth}
        \centering
        \includegraphics[width=\linewidth]{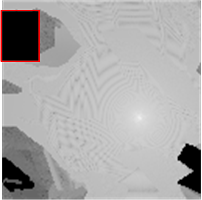}
        \subcaption{inpaint}
    \end{minipage}
    \hfill
    \begin{minipage}{0.125\textwidth}
        \centering
        \includegraphics[width=\linewidth]{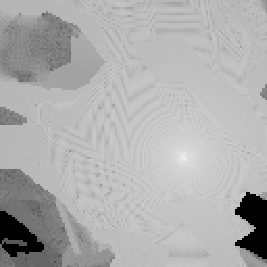}
        \subcaption{\textbf{proposed}}
    \end{minipage}
    \hfill
    \begin{minipage}{0.125\textwidth}
        \centering
        \includegraphics[width=\linewidth]{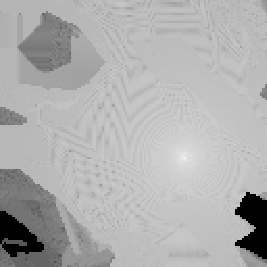}
        \subcaption{KNN}
    \end{minipage}
    \hfill
    \begin{minipage}{0.125\textwidth}
        \centering
        \includegraphics[width=\linewidth]{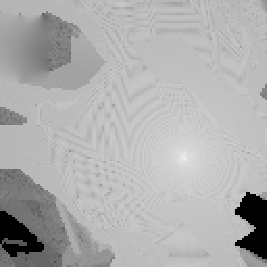}
        \subcaption{Kriging}
    \end{minipage}
    \hfill
    \begin{minipage}{0.125\textwidth}
        \centering
        \includegraphics[width=\linewidth]{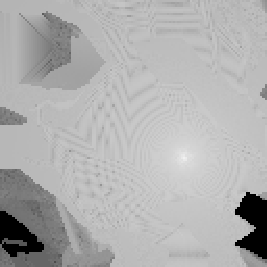}
        \subcaption{Bilinear}
    \end{minipage}
    \hfill
    \begin{minipage}{0.125\textwidth}
        \centering
        \includegraphics[width=\linewidth]{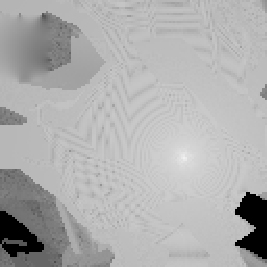}
        \subcaption{RBF}
    \end{minipage}
    
    % \caption{Visualization results of different methods for CKMImageNet dataset when mask not convering buildings. (a) Ground truth. (b) Masked image. (c) Diffusion model. (d) KNN. (e) Kriging. (f) Bilinear. (g) RBF. }
    \caption{Visualization results of different methods without masking buildings.}
    \label{fig:ckm_seven_images_withoutbuliding}
\end{figure*}

Fig.~\ref{fig:ckm_seven_images_withoutbuliding} demonstrates the visualization examples of CKM images inpainting task when the buildings are not masked. 
The error metrics comparison of the considered methods to test $500$ CKM images are listed in Table~\ref{ErrorMetricsTable2}. 
The reconstruction error of our proposed scheme is the smallest in comparison with the four benchmarking methods, with a RMSE of $10.7758$ dB and a MAE of $5.1412$ dB. 
Besides, in contrast to Table~\ref{ErrorMetricsTable1}, the completion effect when buildings are not covered is significantly better for all schemes. 
% However, current prediction results are still not accurate enough. One reason is that the image size is strictly limited due to the limited amount of data and limited computing power.
% In other words, although building information is complete, the channel knowledge affected by buildings outside the image area is masked, which is plotted as Fig. \ref{fig:ckm_seven_images_withoutbuliding}. 
% In this situation, the model cannot learn the relationship between the channel knowledge and the buildings outside the image area.

\begin{table}[!t]
    \renewcommand{\arraystretch}{1.5}  % 增加行间距
    \caption{Performance Comparison Without Masking Buildings}
    \centering
    \label{ErrorMetricsTable2}
    \setlength{\tabcolsep}{4pt}
    \begin{tabular}{c|c|c|c|c|c}
        \hline
        % \multicolumn{1}{c|}{} & \multicolumn{5}{c|}{\textbf{CKMImageNet}} & \multicolumn{5}{c}{\textbf{RadioMapSeer}} \\ \hline
        % \diagbox{\textbf{Metric}}{\textbf{Method}} & \textbf{KNN} & \textbf{Kriging} & \textbf{Bilinear} & \textbf{RBF} & \textbf{Diff} \\ \hline
        \multicolumn{1}{c|}{} & \textbf{KNN} & \textbf{Kriging} & \textbf{Bilinear} & \textbf{RBF} & \textbf{Proposed} \\ \hline
        \textbf{MSE$({\rm{d}}{{\rm{B}}^2})$}  & $245.7755$ & $272.5737$ & $474.6037$ & $188.2536$ & \textbf{116.1187} \\ \hline
        \textbf{NMSE}  & $0.0041$ & $0.0046$ & $ 0.0079$ & $0.0032$ & \textbf{0.0019}  \\ \hline
        \textbf{RMSE$({\rm{dB}})$}  & $15.6772$ & $16.5098$ & $21.7854$ & $13.7206$ & \textbf{10.7758}  \\ \hline
        \textbf{MAE$({\rm{dB}})$}   & $5.8437$ & $8.5845$ & $8.0328$ & $6.0715$ & \textbf{5.1412} \\ \hline
    \end{tabular}
\end{table}

% \begin{table*}[!t]
%     \renewcommand{\arraystretch}{1.5}  % 增加行间距
%     \caption{Performance Comparison when masking the BS for two datasets }
%     \centering
%     \label{ErrorMetricsTable}
%     \begin{tabular}{c|c|c|c|c|c|c|c|c|c|c}
%         \hline
%         \multicolumn{1}{c|}{} & \multicolumn{5}{c|}{\textbf{CKMImageNet}} & \multicolumn{5}{c}{\textbf{RadioMapSeer}} \\ \hline
%         \diagbox{\textbf{Metric}}{\textbf{Method}} & \textbf{KNN} & \textbf{Kriging} & \textbf{Bilinear} & \textbf{RBF} & \textbf{Proposed} & \textbf{KNN} & \textbf{Kriging} & \textbf{Bilinear} & \textbf{RBF} & \textbf{Proposed}\\ \hline
%         \textbf{MSE$({\rm{d}}{{\rm{B}}^2})$}   & $42.4357$ & $825.4937$ & $45.5672$ & $212.3440$ & \textbf{27.3767} & $42.2347$ & $200.2108$ & $32.5619$ & $ 36.2842$ & \textbf{3.1976} \\ \hline
%         \textbf{NMSE}  & $0.0007$ & $0.0138$ & $0.0008$ & $0.0035$ & \textbf{0.0005} & $0.0020$ & $0.0094$ & $0.0015$ & $0.0017$ & \textbf{0.0001} \\ \hline
%         \textbf{RMSE$({\rm{dB}})$}  & $6.5143$ & $28.7314$ & $6.7503$ & $14.5720$ & \textbf{5.2323} & $6.4988$ & $14.1496$ & $5.7063$ & $6.0236$ & \textbf{1.7882} \\ \hline
%         \textbf{MAE$({\rm{dB}})$}   & $2.6440$ & $13.0177$ & $ 2.6920$ & $4.6491$ & \textbf{2.1114} & $1.5901$ & $4.6919$ & $1.4657$ & $1.4134$ & \textbf{0.2524} \\ \hline
%         \textbf{BS localization error$({\rm{m}})$}   & $8.81$ & $9.12$ & $8.88$ & $13.71$ & \textbf{3.30} & $6.44$ & $6.32$ & $6.56$ & $6.40$ & \textbf{2.00} \\ \hline
%     \end{tabular}
% \end{table*}

\begin{table}[!t]
    \renewcommand{\arraystretch}{1.5}  % 增加行间距
    \caption{Performance Comparison When Masking BS}
    \centering
    \label{ErrorMetricsTable}
    \setlength{\tabcolsep}{4pt}
    \begin{tabular}{c|c|c|c|c|c}
        \hline
        % \multicolumn{1}{c|}{} & \multicolumn{5}{c|}{\textbf{CKMImageNet}} & \multicolumn{5}{c}{\textbf{RadioMapSeer}} \\ \hline
        \multicolumn{1}{c|}{} & \textbf{KNN} & \textbf{Kriging} & \textbf{Bilinear} & \textbf{RBF} & \textbf{Proposed} \\ \hline
        \textbf{MSE$({\rm{d}}{{\rm{B}}^2})$}   & $42.4357$ & $825.4937$ & $45.5672$ & $212.3440$ & \textbf{27.3767} \\ \hline
        \textbf{NMSE}  & $0.0007$ & $0.0138$ & $0.0008$ & $0.0035$ & \textbf{0.0005} \\ \hline
        \textbf{RMSE$({\rm{dB}})$}  & $6.5143$ & $28.7314$ & $6.7503$ & $14.5720$ & \textbf{5.2323} \\ \hline
        \textbf{MAE$({\rm{dB}})$}   & $2.6440$ & $13.0177$ & $ 2.6920$ & $4.6491$ & \textbf{2.1114} \\ \hline
        \textbf{BS loc. err.$({\rm{m}})$}   & $8.81$ & $9.12$ & $8.88$ & $13.71$ & \textbf{3.30} \\ \hline
    \end{tabular}
\end{table}

\begin{figure*}[htbp]  % 使用 figure* 环境以跨越两栏
    \centering
    \begin{minipage}{0.125\textwidth}  % 每个子图宽度
        \centering
        \includegraphics[width=\linewidth]{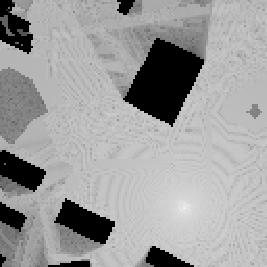}  % 替换为你的图片路径
        \subcaption{ground truth}
    \end{minipage}
    \hfill  % 水平填充以分隔子图
    \begin{minipage}{0.125\textwidth}
        \centering
        \includegraphics[width=\linewidth]{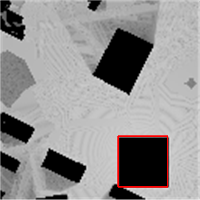}
        \subcaption{inpaint}
    \end{minipage}
    \hfill
    \begin{minipage}{0.125\textwidth}
        \centering
        \includegraphics[width=\linewidth]{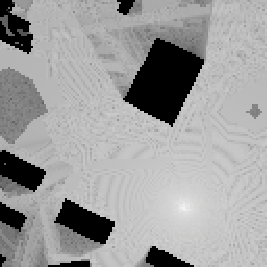}
        \subcaption{\textbf{proposed}}
    \end{minipage}
    \hfill
    \begin{minipage}{0.125\textwidth}
        \centering
        \includegraphics[width=\linewidth]{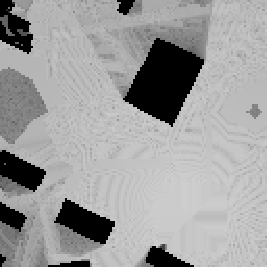}
        \subcaption{KNN}
    \end{minipage}
    \hfill
    \begin{minipage}{0.125\textwidth}
        \centering
        \includegraphics[width=\linewidth]{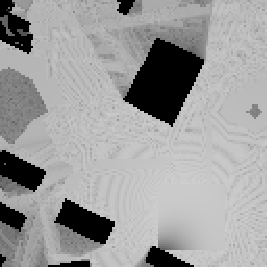}
        \subcaption{Kriging}
    \end{minipage}
    \hfill
    \begin{minipage}{0.125\textwidth}
        \centering
        \includegraphics[width=\linewidth]{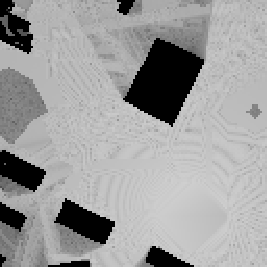}
        \subcaption{Bilinear}
    \end{minipage}
    \hfill
    \begin{minipage}{0.125\textwidth}
        \centering
        \includegraphics[width=\linewidth]{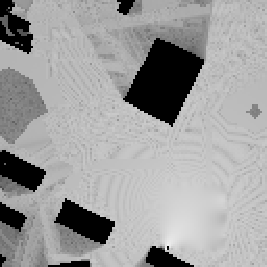}
        \subcaption{RBF}
    \end{minipage}
    
    % \caption{Visualization results of different methods for CKMImageNet dataset. (a) Ground truth. (b) Masked image. (c) Diffusion model. (d) KNN. (e) Kriging. (f) Bilinear. (g) RBF. }
    \caption{Visualization results of different methods to localize the BS.}
    \label{fig:ckm_seven_images}
\end{figure*}

It is worth mentioning that our proposed approach is also quite effective for base station (BS) localization in complex environment. 
% Specifically, we consider the missing channel knowledge of the ${\rm{64m}} \times {\rm{64m}}$ area including the BS, within the ${\rm{256m}} \times {\rm{256m}}$ range in the real environment, 
Specifically, we consider the channel knowledge of the ${\rm{64m}} \times {\rm{64m}}$ area including the BS in the real physical environment within the range of ${\rm{256m}} \times {\rm{256m}}$ is missing, 
whose visualization effects of different methods for CKMImageNet dataset is shown in Fig.~\ref{fig:ckm_seven_images}.
% We use the diffusion model and several interpolation methods to inpaint the CKM images with missing BS information, as shown in Fig. \ref{fig:ckm_seven_images}. 
% In addition, to further demonstrate the effectiveness of the algorithm, we also complete the CKM images of the RadioMapSeer dataset which has a size of 40000 $256 \times 256$ images and visualization results are plotted in Fig. \ref{fig:rm_seven_images}.
% Table. \ref{ErrorMetricsTable} lists the error indicators in predicting channel knowledge of the BS's vicinity and the BS localization error for the CKMImageNet and RadioMapSeer datasets for five schemes, where the testing dataset size is $500$.
Table~\ref{ErrorMetricsTable} lists the error indicators in predicting channel knowledge of the BS's vicinity and the BS localization error for five schemes, where the testing dataset size is $1000$.
Note that the diffusion model is much better than the interpolation methods in predicting the channel knowledge near the BS and the positioning performance of the BS. 
% For CKMImageNet, the MSE can reach ${\rm{27}}{\rm{.3767d}}{{\rm{B}}^2}$. 
Compared with the BS position in the ground truth, the positioning error is only ${\rm{3}}{\rm{.30m}}$ even in complex urban environment without LoS links in most cases. 
% In addition, one reason why the RadioMapSeer dataset has better inpainting performance is that the channel propagation characteristics that need to be learned from this dataset are simpler. 
% In contrast, the building heights of the CKMImageNet dataset are from the real values provided by OpenStreetMap, and the ground is not flat. Therefore, the distribution of CKMIMagenet's channel knowledge is more complex but closer to reality.

\section{Conclusion}
% This paper proposes to utilize generated diffusion model to restore CKM images with limited channel knowledge data.
% We first use the region-specific channel knowledge datasets to generate new channel knowledge maps based on the basic diffusion model, and then obtain pretrained model.
% Furthermore, we use pretrained model to solve the inverse problem for diffusion models based on the incomplete channel knowledge database.
% Simulation results have illustrated channel knowledge maps which are generated from Gaussian noise. Later, we will continue to study how to use inverse problem to solve CKM image reconstruction.
In this paper, we proposed a novel generative CKM construction method based on conditional diffusion models with partially observed channel knowledge data. 
Simulation results demonstrate the effectiveness of the proposed method across two different scenarios: masking buildings and not masking buildings. 
% We , and the results show that our approach outperforms other benchmarks, achieving the best performance in terms of MSE and other error metrics. 
Additionally, our proposed method for BS localization achieves superior performance on the CKMImageNet dataset, outperforming benchmark methods.

% \small
\bibliographystyle{IEEEtran}
\begin{footnotesize}
\bibliography{reff}
\end{footnotesize}

% \begin{flushleft}
% \fontsize{9pt}{11pt}\selectfont  % 设置字体大小为9pt，行距为11pt
% \bibliography{reff}
% \end{flushleft}

\end{document}